# Verification &Validation of Agent Based Simulations using the VOMAS (Virtual Overlay Multi-agent System) approach

Muaz A. Niazi, Amir Hussain and Mario Kolberg

*Abstract*—Agent Based Models are very popular in a number of different areas. For example, they have been used in a range of domains ranging from modeling of tumor growth, immune systems, molecules to models of social networks, crowds and computer and mobile self-organizing networks. One reason for their success is their intuitiveness and similarity to human cognition. However, with this power of abstraction, in spite of being easily applicable to such a wide number of domains, it is hard to validate agent-based models. In addition, building valid and credible simulations is not just a challenging task but also a crucial exercise to ensure that what we are modeling is, at some level of abstraction, a model of our conceptual system; the system that we have in mind. In this paper, we address this important area of validation of agent based models by presenting a novel technique which has broad applicability and can be applied to all kinds of agent-based models. We present a framework, where a virtual overlay multi-agent system can be used to validate simulation models. In addition, since agent-based models have been typically growing, in parallel, in multiple domains, to cater for all of these, we present a new single validation technique applicable to all agent based models. Our technique, which allows for the validation of agent based simulations uses VOMAS: a Virtual Overlay Multi-agent System. This overlay multi-agent system can comprise various types of agents, which form an overlay on top of the agent based simulation model that needs to be validated. Other than being able to watch and log, each of these agents contains clearly defined constraints, which, if violated, can be logged in real time. To demonstrate its effectiveness, we show its broad applicability in a wide variety of simulation models ranging from social sciences to computer networks in spatial and non-spatial conceptual models.

## I. INTRODUCTION

VALIDATION of any simulation model is a crucial task[1, 2]. Simulations, however well-designed, are always only an approximation of the system and if it was so easy to build the actual system, the simulation approach would never have been used [3]. Of all the simulation models, agent-based modeling and simulation paradigm has recently gained a lot of popularity by being applied to a very wide range of domains such as [4-9]. Validation of models typically requires experts to look at data or animation as errors and un-wanted artifacts can appear in the development of agent-based models [10]. However, because of the complex nature of agent-based models comprising of multiple interacting entities and the strong dynamics and frequent emergence patterns in the system, it can be hard to validate agent-based models in the same way as traditional simulation models.

In the case of agent-based simulations, it is even easier to fall into the trap of tweaking the variables, especially since occasionally, the inputs can tend to be quite numerous [11]. Because of the complex nature of agent based models and resulting emergence as shown in [12-14], coupled with an enormous variation possibility of the variables, the results of the simulation study can vary considerably by changing the range or even the step size of just one or two variables. Thus, it is vitally important to be able to validate the agent-based simulation. The problem however, comes from the grounds up since validation is not to be an after-thought; it needs to be initiated alongside at the start of the simulation study. Now, validation of agent based models can be quite a challenging task [15, 16]. One problem lies in the fact that validation typically requires SME (Subject Matter Experts) to analyze [3] the simulation data or animation for comparison with another system or model. However, because of appearance of complex phenomenon such as emergence of behavior, where one plus one is not necessarily two as it depends more on the two "ones" and the behavior of the addition operation as is the norm in complex systems as compared to complicated systems [17]. Thus it can be very difficult to be sure if the behavior that we are observing is truly representative of the actual system[18]. Also, it is important to note here that even models, which cannot be validated might have merit and use such as bookkeeping devices or as an aid in selling ideas or as a training aid or even as part of an automatic management system. In the social sciences literature and ACE (Agents in Computation Economics), empirical validation of agent-based models has been described in [19]. Alternate approaches to empirical validation are discussed in [20]. Replication of agent-based models has been considered very important by some authors and has been discussed in [21]. An approach of validation based on philosophical truth theories in simulations has been discussed in [22]. Another approach called "companion modeling" is an iterative participatory approach where multidisciplinary researchers and stakeholders work together continuously throughout a four-stage cycle: field study and data analysis; role-playing games; agent-based model design and implementation; and intensive computational experiments [23]. Agent-based social simulation has also been used for validation and calibration [24].

In the past, although agent-based simulation has been shown to be useful in the validation of multi-agent systems[25, 26], multi-agent systems have not been used to validate agent-based models. On the other hand, simulations have been used in conjunction with software engineering for a long time[27]. Our work can be considered as pertaining to the last two stages of "Companion Modeling" i.e. Agent-Based Model Design/Implementation as well as Intensive Computational Experiments. Specifically, in this paper, we present the following innovations:

- We show how to develop a VOMAS (Virtual Overlay Multi-Agent System), which can be used for the validation of agent based simulation models.
- We thus further develop social science based validation techniques that can be applicable to both social science as well as other relevant domains.
- We present an object-oriented software engineering based methodology for validation of agent-base models, which provides for both logging as well as animation based validation approaches in addition to test-case/invariant based approaches.

The rest of the paper is structured as following: First we give an overview of the terms "Verification", "Validation" and "Credibility" as discussed in the literature. We also discuss how these terms have been considered traditionally in simulation models. Next, we give an overview of performing Validation using VOMAS. We show the design of VO (Virtual Overlay) and Logger agents. Next, we show an example of developing a VOMAS for an existing model from Agent-Based Modeling literature, and demonstrate its usefulness, and ease in validation. Finally we conclude the paper.

## II. VERIFICATION, VALIDATION AND CREDIBILITY

Researchers transform real-world systems to models by applying abstraction. This transformation requires propagating concepts from the real world to useful computational models. These, in turn, are used to develop simulations. Simulation models, in essence end up giving back results which can be useful for the real world. As such, the more effective the abstraction mechanism, the better would be the expected real world benefits.

### A. Peculiarities of Agent-Based Models

In case of agent-based models, the simulation comprises of one or more agents. These agents can work independently or else interact with each other. These computational entities, which are typically, simplifications of real-world counter-parts, need to have some meaningful semantics which can include anywhere from simple behaviors as well as variables for storing different items, such as states, to complex representations such as artificial neural networks, artificial immune systems, cognitive models etc.

### B. Definitions of the terms:

Validation is the process by which we can determine if the model is a representation of the system.[3]. This is always performed while keeping the specific abstraction by the designer in mind. Verification is basically the debugging of the system where we ensure that the model that we build is working correctly. Credibility is achieved when the decision-makers and other key project personnel accept the model as well as its results as "correct".

### C. Correlation with VOMAS?

VOMAS approach has been designed to cater for all kind of agent-based models. As such, it has capability to monitor spatial as well as non-spatial concepts in agent-based models.

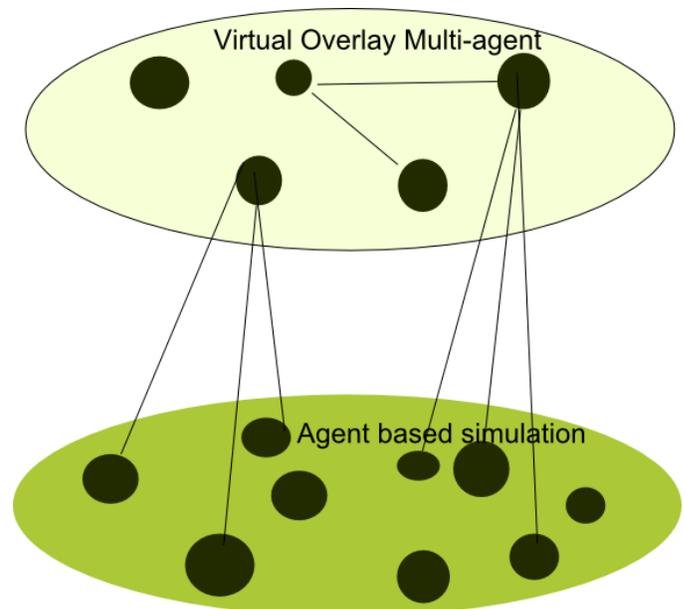

Fig. 1 VOMAS relation with an Agent Based Model

### D. Verification & Validation of agent-based models

One sure way to establish the validity of agent-based model is to have Subject Matter Experts, who give the specification as well as examine the results and logs of simulation runs. VOMAS approach allows experts to be involved in the design of the agent-based model as well as the custom-built VOMAS from scratch. By involving SMEs from the start of the project, which are essentially equivalent to clients in the software engineering domain, VOMAS approach allows the simulation study to be a stronger candidate for success.

## III. VALIDATION USING VOMAS

### A. Validation in agent based simulations

To understand VOMAS, let us examine figure 1. The Virtual Overlay Multi-agent System is created for each simulation model separately by a discussion between the simulation specialist as well as the SMEs (Subject Matter Expert). When the actual simulation is executed, the VOMAS agents perform monitoring as well as logging tasks and can even validate constraints given by the system designer at design time.

*B. A Taxonomy of Agent-Based Validation techniques using VOMAS*

Now, let us examine how agent-based models are structured. Since agent based models have one or more agents, what these agents really mean in the real-world is entirely up to the designer of the simulation. These elements can be spatial in nature, where distance between agents in the simulation is important or else non-spatial, where there is no concept of distance in the simulation as shown in Fig. 2. In case of spatial models, it is also entirely possible that the exact distance may not be important, but the links between agents could be important. An example of this is HIV based models, where interaction between agents can be shown as links.
A detailed description of each of these follows.

Fig. 2 A Taxonomy of Agent Based Validation techniques

1. Visual Validation:
   Visual validation is a face validation technique based on an animation based validation technique where the SME can examine the animation to see if the behavior appears to be similar to that expected in the actual domain.

2. Validation using VOMAS:
   In case of VOMAS, we can validate both spatially as well as non-spatially.

3. Spatial Validation:
   In spatial validation, the placement of agents in the simulation is important. This includes the placement of some of the VOMAS agents, which interact with the actual agent based simulation.

4. Non-Spatial Validation:
   In non-spatial validation, the actual distance is not important. These could be used to validate for aggregate data and constraints/invariants etc.

5. Networked or Link-Based Validation
   In spatial validation, it is possible that the actual placement is less important than the links between them. In case of social simulation, the example could be links to show social network friendships. In case of computer science based networks, these could represent e.g. Connectivity of Peer-to-Peer overlay networks.

6. Proximity Based Validation
   In this case, the actual proximity of agents to each other and especially to VOMAS agents is important. An example of this is pred-prey models where VOMAS agents can verify certain characteristics of agents passing by them at a certain time.

7. Log based validation:
   In log based validation, the SME can specify what things to be watched and logged so that they can be examined after the fact and see how e.g. the populations evolved over time, or else how wireless sensor networks lost their power over time etc.

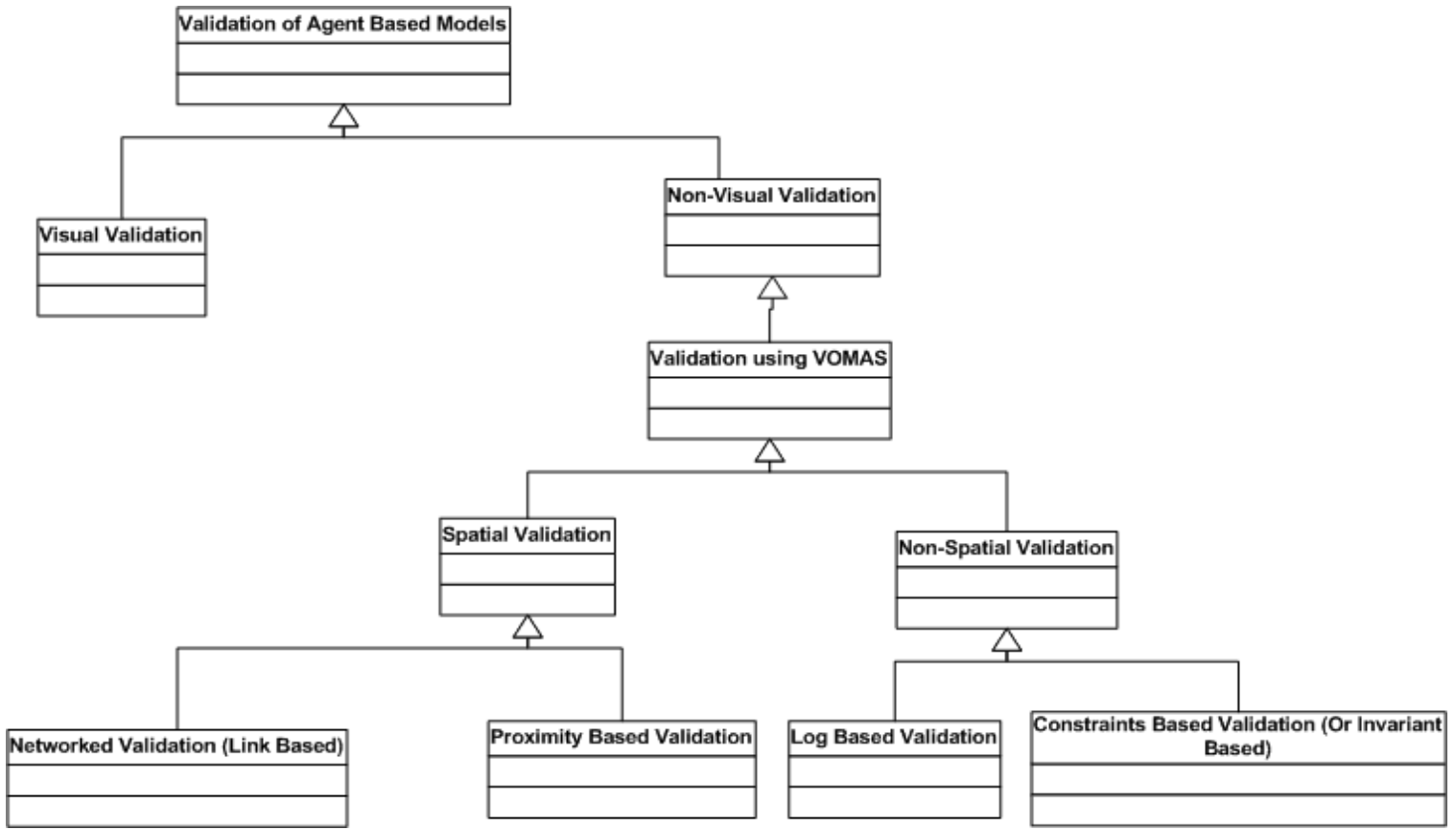

8. Constraint-based validation or Invariant Based Validation:

   It is entirely possible that the SME says that there are certain constraints, which should never be violated in a certain simulation experiment. If these were ever to be violated, then the simulation system should notify the user via some console or else log the event as a special case. E.g. Wolves must never all die in a wolf-sheep predation. If all of the wolves die, then the simulation needs to be stopped etc. as further data collection exercise might not be useful.

C. *Analysis of VOMAS*

The analysis of VOMAS has been conducted based on a scenario-modeling approach. In figure 3, we see the use cases, some of which are described below. The rest should be self-explanatory and we are not listing them for shortage of space:

1) *Verify the Model*

   The SME verifies the model by means of execution of the simulations by the Simulation Specialist. The detailed verification (debugging) is checked by the simulation specialist but in case of any ambiguity, the SME can be referred.

2) *Validate the Model*

   This validation is done in three ways
   a. Validation using animations:
      This validation is face validation by the SME by means of analyzing the animations.
   b. Validation using Logs
      In this case, logs are generated based on watches specified by the SME. These logs show after the fact, the entire scenarios like black boxes from airplanes.
   c. Validation using Invariants
      These can be cases where the SME wants either immediate feedback even while running large scale parameter sweeps. So, if the invariants or constraints are ever violated, the user can be notified. Or at least, this is definitely logged in the simulation log.

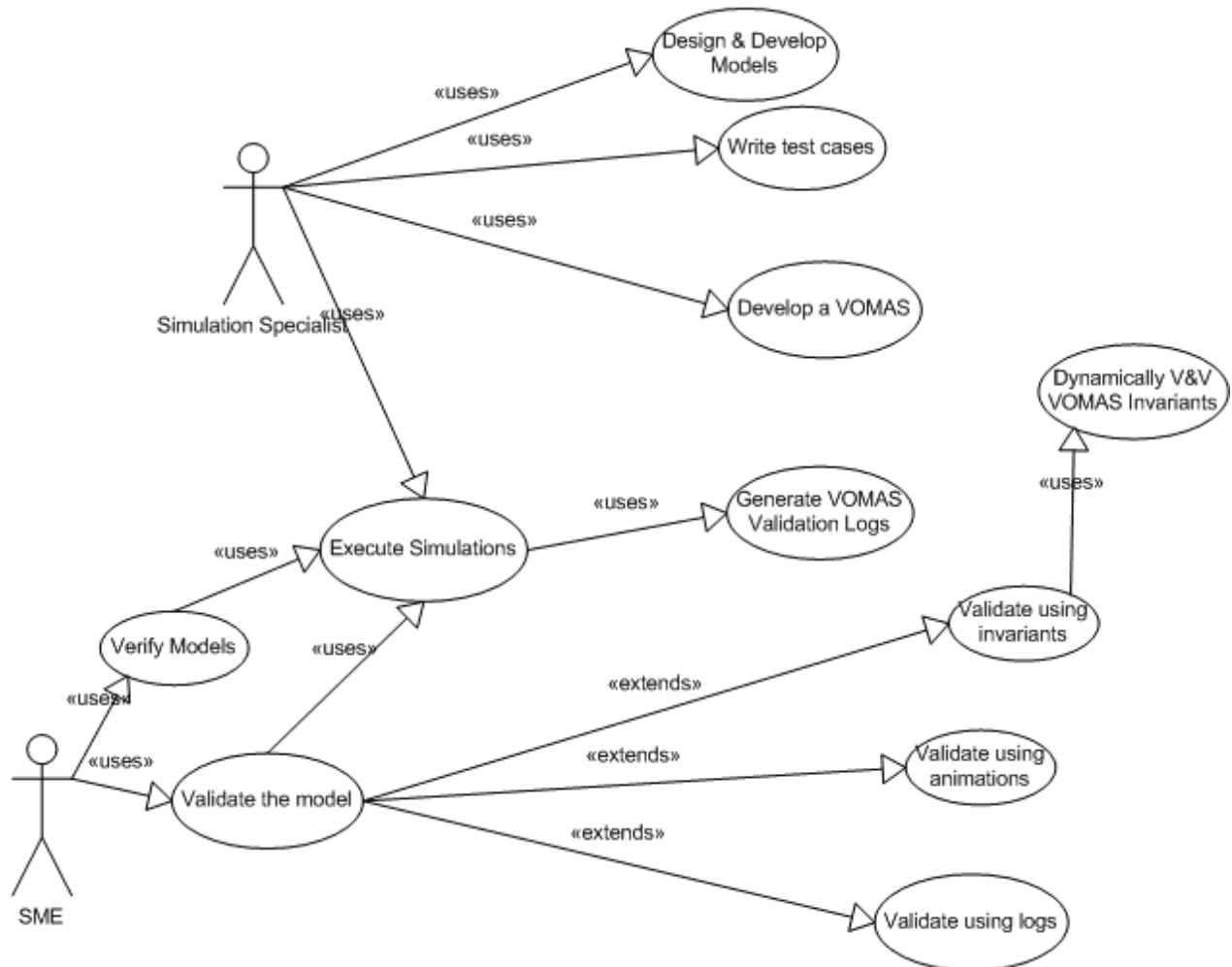

Fig. 3 Use case model of simulation model design and V&V

### 3) Design and Develop Models

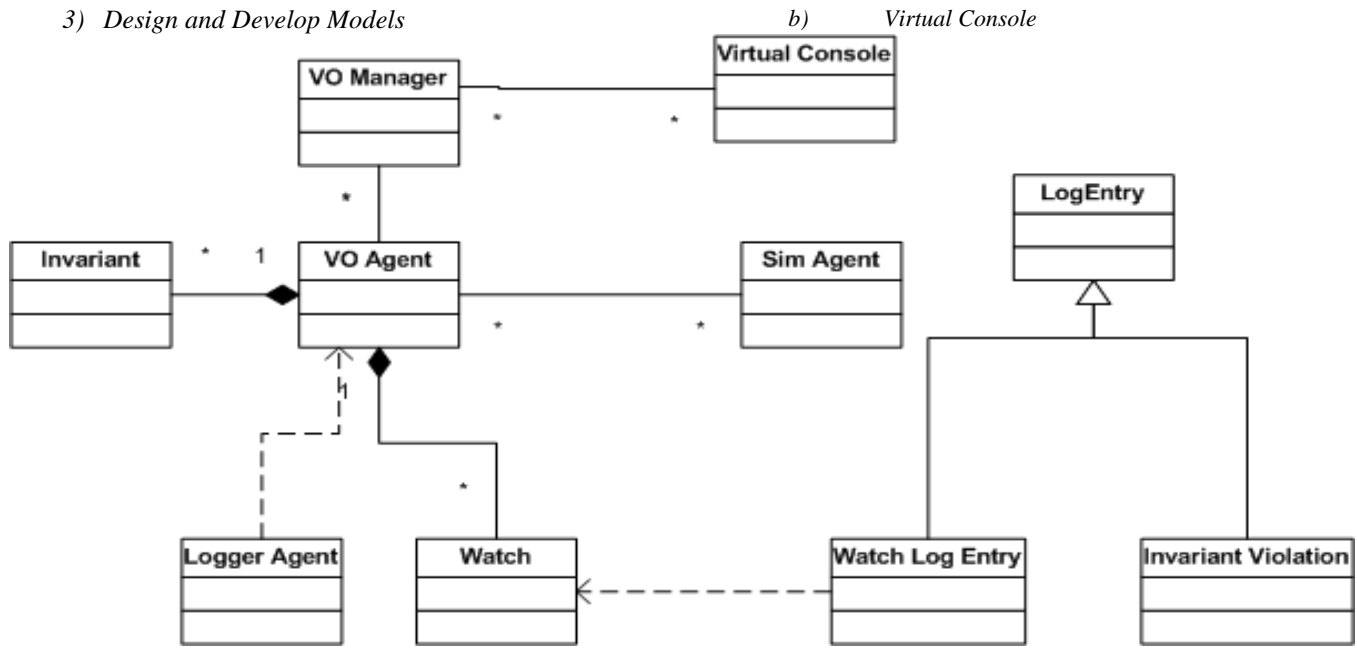

Fig. 4 Class diagram of agents in a VOMAS

This use case is to be conducted by the simulation specialist in conjunction with the SME.

### D. Design of VOMAS

#### 1) Motivation

One of the most popular approaches in Validation is the three step approach given in [28]. The approach has the following steps:

  a)   Build a model that has high face validity.

  b)   Validate model assumptions

  c)   Compare the model input-out transformations to corresponding input-output transformations for the real system.

VOMAS has been designed to cater for both face validity as well as model assumptions and io-transformations. Model assumptions are ensured by the use of invariants. Face validation is ensured by means of various techniques based on spatial and non-spatial validation and animation-based validation. IO-transformations are ensured by means of essential logging components. Thus, in other words VOMAS provides the complete validation package.

#### 2) Description of Class Diagram

In figure 4, we see the class diagram of the VOMAS agents and how they interact with the agents in the simulation. The description of each of these agents is given below:

  a)   VO Manager

VO manager agent is the key agent handling the interaction of all of the other agents.

  b)   Virtual Console

Virtual Console agent is an agent, which can be used to dynamically display various messages at run-time.

  c)   Invariant

Invariant is any condition, which the designer of the VOMAS and the agent-based simulation, feels that must not be violated during the execution of the simulation. If the Invariant is violated, the violation is logged.

  d)   Logger Agent

The logging capability is provided by the Logger Agent.

  e)   Watch

If the designer of the system wants some value to be observed, it can be made a watch.

  f)   Watch Log Entry

Each watch can also be logged as a logged entry.

  g)   Invariant Violation

Invariant violations can be logged at run-time to the Console Virtual agent or else the log as a log entry.

  h)   Log Entry

The base class of all log entries.

  i)   Sim Agent

This is an agent which is part of the agent based simulation model.

  *j)*   *VO Agent*

These are agents which can be located spatially or non-spatially to monitor the entire simulation.

## IV. CASE STUDY

Here, we present application of a VOMAS to an agent-based simulation mode of the "Simulation of the research process". Recently an agent-based simulation model of researchers attempting to present research in International publication venues was presented in [5]. We demonstrate how to develop and use the associated VOMAS on this model.

### A. The Publishing Researchers' model

In the publishing researcher model, the abstraction is that researchers are modeled as agents in the simulation. The higher the publications of an agent, the higher the agent goes. Thus space in this simulation model is essentially used to show the capability of the researcher. A screenshot of the simulation model is shown in fig 5. For more details, the interested reader is advised to consult the original article. The model has been developed using NetLogo [29]. So, let us formally define some of the entities involved:

*SME*: An Expert Researcher with experience of publishing in various venues.

*Objective of Simulation Study*: To examine how the policies of researchers in selection of publication venues impacts an overall organization.

Example Invariant:
Basis: In a particular simulation experiment, enough time of simulation run should be given to ensure that journal preferring researchers publish at least ten times during the simulation.
Invariant: If simulation stops before each journal preferring researcher is able to publish at least ten times, note an invariant violation in the console and/or the log.

Example watches:
Measure the total number of researchers with the best policy.
Measure the number of researchers above a certain threshold.
Measure the number of overall publications.

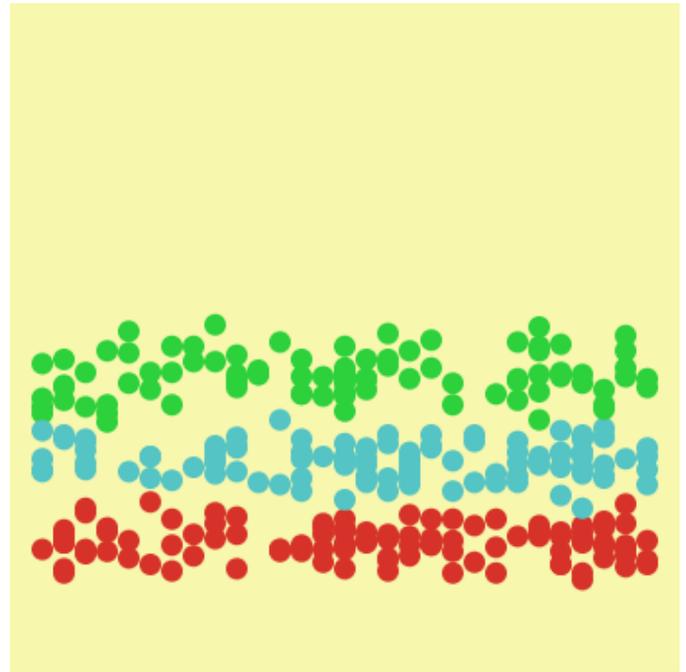

Fig. 5 Screenshot of the researchers' model [5] showing researchers according to their publication count. (Lime = Conference preferring, Red = Journal Preferring, Cyan = No Preference)

## V. CONCLUSION AND FUTURE WORK

In this paper, we have presented a novel framework for the validation of agent based simulation models. We have given a description of how VOMAS agents can be constructed for validation. As a case study, we have shown its application on an existing published model. In the future, we shall apply VOMAS on various types of simulation models and demonstrate how it can be effective in validation. Some of the models we intend to explore VOMAS application on, include pred-prey models, tumor growth models, Peer-to-Peer unstructured overlay network models.